\documentclass[11pt,conference]{IEEEtran}
\usepackage{amssymb}
\newtheorem{thm}{Theorem}
\newtheorem{lem}[thm]{Lemma}

\newcommand{\F}{\mathbb{F}}
\newcommand{\C}{\mathcal{C}}

\begin{document}

% paper title
\title{Coding Solutions for\\ the Secure Biometric Storage
  Problem\footnotemark{$^{\scriptscriptstyle{1}}$}}

\author{ \authorblockN{Davide Schipani\footnote{TEST}}
  \authorblockA{University of Z\"urich \\
    Mathematics Institute \\
    CH-8057 Zurich,  Switzerland 
    } \and \authorblockN{Joachim Rosenthal}
  \authorblockA{University of Z\"urich \\
    Mathematics Institute \\
   CH-8057 Zurich,  Switzerland 
    }\footnote{Test} }

\maketitle
\footnotetext[1]{Authors were supported in part by Swiss National Science Foundation Grant no.\
    107887 and no.\ 113251. A first version of this paper was
    published in the arXiv in January 2007.}

% The {\bf Abstract} section should be no more than $250$ words and
% should contain no math notation. If citations are required in the abstract,
% they should be self-contained, e.g. Shannon, \emph{Bell Syst.\ Tech.\ J.} 1948,
% rather than [1]. The abstract will be published separately in the hardcopy book
% of abstracts.

\begin{abstract}
 
  The paper studies the problem of securely storing biometric
  passwords, such as fingerprints and irises. With the help of
  coding theory Juels and Wattenberg derived in 1999 a scheme
  where similar input strings will be accepted as the same
  biometric. In the same time nothing could be learned from the
  stored data. They called their scheme a \emph{fuzzy commitment
    scheme}. 
  
  In this paper we will revisit the solution of Juels and
  Wattenberg and we will provide answers to two important
  questions: What type of error-correcting codes should be used
  and what happens if biometric templates are not uniformly
  distributed, i.e. the biometric data come with redundancy.
  
  Answering the first question will lead us to the search for 
  low-rate large-minimum distance error-correcting codes which
  come with efficient decoding algorithms up to the designed
  distance. 
  
  In order to answer the second question we relate the rate required with a quantity connected to the
  \lq\lq{}entropy'' of the string, trying to estimate a sort of \lq\lq{}capacity'',
  if we want to see a flavor of the converse of Shannon's noisy
  coding theorem.
  
  Finally we deal with side-problems arising in a practical
  implementation and we propose a possible solution to the main one that
  seems to have so far prevented real life applications of the
  fuzzy scheme, as far as we know.
\end{abstract}

\section{Introduction}

Traditionally passwords for access to a computer are not stored
in plain-text but rather as images under a hash function. Hash
functions have the property that they can easily be computed for
any input string but it is computationally not feasible to
compute any pre-image of a given image point. Usually it is also
desirable that hash functions are `collision resistant', this
means it is computationally not feasible to come up with two
different input strings which are mapped to the same hash
values. Because of the last property standard hash functions such
as SHA-1 are not suitable to store biometric data. What we would
need is a hash function having the property that similar input
strings will result in the same hash values. Until recently no
good scheme has been known and many practical systems store
biometric data such as fingerprints to access a personal computer
in plain-text.

Martinian, Yekhanin and  Yedidia~\cite{ma05c} call the problem 
at hand \emph{the secure biometric storage problem}. The problem
arises  when biometrics such as fingerprints and irises are used
instead of passwords. It is desirable for  security reasons that
the biometric data is not stored in plain-text on a storage
device but rather in encrypted form. When a user wants to access
the system the access device should grant access as long as two
biometrics do not differ by more than a certain amount of bits. 

In the literature there are several schemes which use ideas from
coding theory to tackle the secure biometric storage problem.
According to the authors of~\cite{ma05c} the first solution was
proposed by Davida, Frankel and Matt in~\cite{da98p}. In their
own paper~\cite{ma05c} Martinian et. al. propose an information
theoretic solution based on the Slepian-Wolf theorem. This
system has the property that the biometric is securely stored, it
has however the disadvantage that a person who has access to the
stored data and the implemented algorithm can compute a bit
string which will provide access to the system even though the
bit string is not close to any biometric data.

In this paper we will be concerned with an algorithm first
proposed by  Juels and  Wattenberg~\cite{ju99}. Also this system
makes heavily use of coding theory.

The paper is structured as follows: In the next section we
revisit the algorithm of  Juels and  Wattenberg. The original
paper~\cite{ju99} leaves two important questions open. First what
are good practical codes to be used having very large block
length and which provide the robustness and security level
required for the secure biometric storage problem. We provide
answers to this problem in Section~III. The second
question arises when the possible biometric bit-strings are not
uniformly distributed. Of course this is an important issue as
all practical systems are suffering this problem. We will address this problem in Section~IV.

%%%%%%%%%%%%%%%%%%%%%%%%%%%%%%%%%%%%%%%%%%%%%%%%%%%
\section{The Fuzzy Commitment Scheme of Juels and  Wattenberg}

Juels and Wattenberg~\cite{ju99} proposed a `a fuzzy commitment
scheme' capable of storing biometric data in binary form. In this
section we describe the scheme for data over a general alphabet
and we derive a strengthened theorem.

Let $\F=\F_q$ be a finite field. We assume that the biometric
data is given in form of a vector $b\in\F^n$. Assume $\C\subset
\F^n$ is an $[n,k,d]$ linear code and distance $d$ is given by
$$
d=2t+1.
$$
We also assume that there is an efficient decoding algorithm
capable of decoding up to $t$ errors.

Let $h:\F^n\longrightarrow \F^l$ be a hash function. In
particular $h$ should be collision resistant and it should be
computationally not feasible to compute an $x\in h^{-1}(y)$ for
any $y\in\F^l$.
 
Let $b\in\F^n$ be the  biometric one wants to store on the
computer. The algorithm requires to select a random code word
$r_b\in \C$. The system then computes the vector
$$
l:=b-r_b
$$
and stores on the system:
$$
(h(r_b),l).
$$
The following is a strengthening of the main theorem
in~\cite{ju99}.
\begin{thm}                       \label{J-W-Thm}
If the possible biometrics $b\in\F^n$ are uniformly distributed
then computing the biometric $b\in\F^n$ from the stored data
$(h(r_b),l)$ is computationally equivalent to invert the
`restricted' hash function
$$
h\mid_C: C\longrightarrow\F^l.
$$
\end{thm}
\begin{proof}
Since $b$ and $r_b$ were selected independently and uniformly at
random the vector $l:=b-r_b$ reveals no information about the
random choice of $r_b\in\C$. An attacker is left with the task to
compute $r_b$ from $h(r_b)$.
\end{proof}

The theorem provides the means to come up with a practical secure
storage system once we can assume that the biometrics are
uniformly distributed over the ambient space $\F^n$. If this is
the case and if $h$ is a hash function which is practically
secure then we only have to require that the size of the code
$|C|\geq 2^{80}$. This is due to the fact that it is generally
accepted that a total search space of $2^{80}$ is beyond the
capabilities of modern computers. As a result it is desirable
that the constructed codes have dimension $k=\dim\C\geq 80$.\medskip

The following lemma shows that the system allows to accept an
authorized user as soon as this user provides a biometric vector
which comes close enough to the originally supplied vector
$b\in\F^n$.
\begin{lem}
Let $\tilde{b}\in\F^n$ be a vector whose Hamming distance satisfies:
$$
d_H(b,\tilde{b})\leq t.
$$
Then it is possible to efficiently compute $b$ from the stored
data $(h(r_b),l)$. (In fact authorization is granted by comparing the hash stored with the hash of the decoded codeword, without any need to compute $b$.)
\end{lem}
\begin{proof}
$$
d_H(r_b,\tilde{b}-l)=d_H(b-l,\tilde{b}-l)=d_H(b,\tilde{b})\leq t.
$$
The vector $\tilde{b}-l$ decodes by assumption uniquely to the
code vector $r_b$. Knowing $r_b$ and $l$ is equivalent to knowing
$b$. 
\end{proof}

Several considerations are due at this moment, starting with the
choice of the code to use.

In \cite{ju99} it is proposed  that Reed-Solomon and BCH codes might
provide useful results (see also \cite{ha05b}). We believe these
are not necessarily good options for two reasons. First practical biometric
systems have often to deal with large amount of bits (an
estimate in some circumstances could be $10'000$ bits). Moreover we can say an error tolerance of
$10\%$ of errors is a reasonable requirement. BCH codes of block
length $10^4$ and distance $2'000$ are necessarily of very low
rate and it is practically not feasible to run e.g. a Berlekamp-Massey
algorithm once so many syndromes are involved.

The next section addresses the choice of the code.

%%%%%%%%%%%%%%%%%%%%%%%%%%%%%%%%%%%%%%%%%%%%%%%%%%%%%%%%%%%%%
\section{Choice of the code}

Based on the comments in the last section we require an $[n,k,d]$
linear code whose dimension is $k\geq 80$ over the binary field,
possibly smaller if one works over larger alphabets. In addition
one wants to have a large relative minimum distance that only low
rate codes can afford. Indeed because e.g. of the asymptotic
Elias upper bound (see e.g.~\cite{bl03}) only very low rate
binary codes can have relative distance larger than
e.g. $0.4$. Of course the code should come with efficient
decoding algorithms even when the block length is in the range of
$n=10^4$.

We think of two types of codes as possible candidates for this
application, namely 1) {\em Product codes},  and 2) {\em  LDPC
  codes}. Both these codes can be decoded with linear or close to
linear complexity in the block length.

Let us consider the first option: product of classical codes. We
can define them using the generator matrices (see e.g.~\cite{ma77}):
If $A$ and $B$ are the generator matrices of two codes, $C_1$ and
$C_2$, with parameters $(n_1,k_1,d_1)$ and $(n_2,k_2,d_2)$, then
the Kronecker product of matrices

\[A\otimes B= (a_{ij}B)\]
obtained by replacing every entry $a_{ij}$ of $A$ by $a_{ij}B$ is
the generator matrix of the product code.

The new code has parameters $(n_1 n_2, k_1 k_2, d_1 d_2)$ and can
be viewed as the set of all codewords consisting of $n_1\times
n_2$ arrays constructed in such a way that every column is a
codeword of the first code and every row is a codeword of the
second one.

Clearly, given the definition of the product of two codes, the
product of more than two codes can be defined as well.

We give here some examples of product of two codes with parameters
getting close to $(100000,100,20000)$:
\begin{itemize}
  
%\item $(256,80,45)$, a classical Goppa code and $(400,1,400)$, a repetition code;

\item $(512,98,93)$, a classical Goppa code and $(200,1,200)$, a repetition code.

\item $(121,49,37)$, an extended Goppa code~\cite{pr01}
  and $(825,2,550)$, where codewords are the all-zero codeword, two codewords
  with respectively the first and the last $275$ bits equal to
  ones and the other zeroes, and the sum of these two;

\item $(144,50,48)$, an extended Goppa code %~\cite{pr01}
  and $(693,2,462)$, where codewords are the all-zero codeword, two codewords
  with respectively the first and the last $231$ bits equal to
  ones and the other zeroes, and the sum of these two;
 
\item $(256,26,116)$, an extended Goppa code %(cfr. \cite{pr01})
  and $(400,4,200)$, an (8,4,4) extended Hamming code with each symbol repeated 50
  times.
  
\end{itemize}

%We give also an example to obtain larger minimum distance: $(400,100,151)$, an extended Goppa code and $(250,1,250)$, a repetition code.

The decoding procedure of such product codes is based on
iterative algorithms, where one decodes alternatively by columns
and by rows (see also \cite{me06b,mo05,mo02b}). Thanks to this
kind of splitting in the decoding, we can afford to use classical
codes such as Goppa codes, while maintaining a reasonable
computational complexity.

Since the first version of our paper was made available at the
arXiv a similar choice of coding scheme was proposed in~\cite{br07}.

As for LDPC codes\nocite{ga62}, the difficulty seems mainly that
of finding the parameters we need. Codes studied in the
literature often aim at rates of 1/2 or higher. Such codes
necessarily have a relatively poor relative minimum distance

Among the many constructions in the literature, we believe that
RA, IRA and eIRA codes (see for example~\cite{ri01a2,ry04})
should be good candidates with this respect. We have also taken
into consideration the use of algebraic constructions of LDPC
codes, such as the Margulis-Ramanujan type~\cite{ro00p}:
in this case we should modify the construction to lower the rate,
for example by taking $m+1$ copies of the graph on the left and
$m$ on the right for a suitable $m$, but we face the difficulty
of finding a good minimum distance~\cite{ma03b}.

Actually turbo codes could be a better option for a low rate; though in more pratical scenarios, as we will see in next section, such low rates are not convenient anymore for security reasons and more standard parameters suit better.

%%%%%%%%%%%%%%%%%%%%%%%%%%%%%%%%%%%%%%%%%%%%%%%%%%%%%%%
\section{Distribution of biometric templates}

Theorem~\ref{J-W-Thm} works under the strong assumption that the
biometric data is uniformly and randomly distributed over the
ambient space $\F^n$. In practical applications this is a very
unlikely scenario. In this section we estimate a threshold for the dimension $k$ of the code, above which the commitment scheme of Juels and
Wattenberg is most probably secure.

First note that if one has some information about the biometric
$b$ it will be possible to recover from $l$ some information
about $r_b$. Dependent on the size of $\C$ it might be possible to do
a search among all codewords with a particular pattern and
consequently break the system. To possibly defend the system from
this attack, one could essentially take a higher rate code (but
at the expense of lowering the minimum distance). So our next
step is to relate the uncertainty or randomness connected to the
string with the dimension required for the code.

Following \cite{br05}, we can speak of the entropy of a binary
string as the $\log$ in base $2$ of the number of possible
strings: so, for example, for a binary string of length $n$,
where each bit is chosen independently and randomly between $0$
and $1$, the entropy is defined to be $n$ and it is measured in
units of information or Shannon bits (see e.g.~\cite{go02}). If
the string is not random, the entropy is the $\log$ of the number
of the so called \emph{typical sequences}; if, for example, each
bit is chosen independently to be $1$ with probability $p$ and
$0$ with probability $1-p$, then the entropy of the string is $n
h(p)$, where $h(p)=-[p\log p+q\log q]$ is the Shannon function.

Now, let $H(b)$ be the entropy of the biometric. If that is $n$,
that means that biometrics are randomly distributed, then we can
afford a code with dimension $k=k_0$ ($k_0=100$, say). When the
distribution is not really random, then the ``number of
possible strings'' is reduced from $2^n$ to $2^{H(b)}$.

So, roughly speaking, it is like the eavesdropper Eve knows the correct
bit at $n-H(b)$ positions, so that if we want her to search
nevertheless among $2^{k_0}$ codewords, then, counting in the worst case over all
possible strings for those positions, we should need
$2^{k_0}\cdot2^{(n-H(b))}$ codewords, i.e. the dimension should
be
\[k\geq k_0+(n-H(b)).\]

Clearly, as said, we are considering a worst case scenario, so that this requirement makes sense for, let's say, reasonable values of the parameters, that is $k_0<<H(b)$; otherwise $k$ could be asked to be even larger than $n$. Essentially our requirement is purposely asking a bit too much than the strict minimum, which though doesn't waste at all in a security concern.

To see the issue from another view point, we can think of a channel, where at one side we have the message $r_b$ and at the other end there's Eve which tries to decode and get $r_b$ from the pair $H(r_b),l$. The converse of Shannon's noisy coding theorem says
that the probability of correct decoding can be bounded as $2^{-n G(R)}$ where
$G(R)$ is a positive function of the rate $R$ for $R>C$. So in some sense we have estimated the capacity of this channel as $\frac{k_0}{n}+1-\frac{H(b)}{n}$.

(For references on information theory, Shannon's noisy coding
theorem and its converse~\cite{co91b,mc04,sh48,sl73,wo78b}.)
  
\section{Practical Implementation Issues}
The fact that, as far as we know, the fuzzy scheme has not found
yet so many real applications in biometric storage, depends not only in
the way of implementing it as we have discussed it so far, but
also in further practical difficulties that make the problem more
complicated than how we stated it.

The main problem to overcome is the fact that the scheme requires
that the two passwords to be compared are prealigned; and the
difficulty consists in aligning with a password that is not in
the clear. There are also some other aspects one has to improve
or fix; for example one has to take into account the possibility
of erasures and unordered collection of biometric features. The
error distribution is also far from uniform in practical schemes.

In the literature
\cite{bo04,bo05,do04,fr01c,ju06,ka05,ul06,ul04,ya04} we can find
a deeper discussion of all these side problems together with
proposals to attack some of them, each of them with its pros and cons. In the following section we propose
another way of dealing with it, i.e.  we propose to use, instead
of biometrics, some particular histograms derived from them that
can capture important features of the images. As a side effect,
since these histograms are also a means of compression, we would
obtain smaller lengths for the passwords to be hashed and also we
wouldn't need to require such a high minimum distance. So looking
for different and more convenient code parameters could be a
relevant consequence.

%%%%%%%%%%%%%%%%%%%%%%%%%%%%%%%%%%%%%%%%%%%%%%%%%%%%%%%%%%
\section{Histograms and Alignment}
What we essentially want to do to solve the pre-alignment problem
is to somehow transform the biometric passwords and store the
output of the transformation. What we first require from this
\lq\lq{}function'' is to be resistant to noise, changes in
illumination and transformations such as translation and
rotation. The literature~\cite{ha00c,ha03p1,ha04a1} indicates
that the so called ``multiresolution histograms'', that are sets
of intensity histograms of an image at multiple image resolution,
satisfy these prerequisites. So they could possibly solve our
problem, but we require another important feature, i.e. we want
the transformation to be one-to-one or at least that not too many
different biometrics give the same output.  Pass and
Zabih~\cite{pa96,pa99b} worked in this direction and introduced
the notions of histogram refinement and joint histograms.  We
believe that some transformation of this kind that encompasses
these features could be a solution to overcome the problem of
alignment. And also new issues would consequently follow: the
size of error tolerance required (that would be much reduced) and
the choice of other suitable code parameters.
   
%%%%%%%%%%%%%%%%%%%%%%%%%%%%%%%%%%%%%%%%%%%%%%%%%%%%%%%%%%%%%%%
\section*{Acknowledgment}
The authors would like to thank Prof. Michele Elia, EE
Department, Politecnico of Torino, and Marco Dominici, XChanges
Visual Effects srl, Milano, for their contributions and
suggestions in developing some ideas presented in the paper.

\bibliography{huge} \bibliographystyle{plain}\end{document}